\input amstex
\documentstyle{amsppt}
\magnification=\magstep 1
\hsize29pc
\vsize42pc
\baselineskip=24truept

\def\tr{\text{tr}\,}
\def\lx{l_x}
\def\U{\Bbb U}

\def\ol{O\left({1\over \lambda}\right)}
\def\[{\left[}
\def\]{\right]}
\def\({\left(}
\def\){\right)}
\def\ltx{\lambda(x-x_0)/ 2}
\def\lxt{\lambda(x_0-x)/ 2}

\def\l {\lambda }
\def\lt{{\lambda\over 2}}
\topmatter
\title  SYMPLECTIC STRUCTURES AND VOLUME ELEMENTS IN THE FUNCTION 
SPACE  FOR THE CUBIC SCHR\"{O}DINGER EQUATION\endtitle
\author K.L.  Vaninsky\endauthor
\affil School of Mathematics, Institute for Advance Study, \\
Princeton, NJ 08940\\
and\\
Department of Mathematics\\
Kansas State University\\
Manhattan, KS 66502
\endaffil
\email vaninsky@math.ias.edu\endemail
\thanks  The work is  supported by NSF grant DMS-9501002\endthanks
\subjclass  \endsubjclass
\abstract 
We consider  various trace formulas for the cubic Schr\"{o}dinger 
equation in the space of infinitely smooth functions subject 
to periodic boundary 
conditions. The formulas relate conventional 
integrals of motion to the periods of some Abelian differentials 
(holomorphic one-forms) on the spectral curve. 
We show that the periods of Abelian differentials are global coordinates
on the moduli space of spectral curves. 
The exterior derivatives of 
the holomorphic one-forms  are the basic and higher symplectic 
structures on the phase space.  We write explicitly these symplectic structures 
in  $QP$ coordinates. 
We compute the  ratio of two symplectic volume elements 
in  the infinite genus limit. 
\endabstract
\toc
\widestnumber\subhead{13.}
\subhead 1.  Introduction. \endsubhead
\subhead 2.  Zero curvature representation.\endsubhead
\subhead 3.  Monodromy matrix. Expansion at infinity. \endsubhead
\subhead 4.  Spectrum. Some products.\endsubhead
\subhead 5.  Floquet solutions. Differential of quasimomentum. \endsubhead
\subhead 6.  Trace formulas. One-gap potentials.\endsubhead
\subhead 7.  Infinite hierarchy of commuting flows. BA functions. \endsubhead
\subhead 8.  Dual BA function. \endsubhead
\subhead 9. Variational identity. \endsubhead
\subhead 10. Periods are moduli.\endsubhead
\subhead 11. Classical symplectic structure.\endsubhead
\subhead 12. Higher symplectic structures.\endsubhead
\subhead 13. Symplectic volume elements.\endsubhead  

\endtoc
 
\endtopmatter
\rightheadtext{ SYMPLECTIC STRUCTURES AND VOLUME ELEMENTS  }
\document

\subhead 1. Introduction\endsubhead 
We consider the cubic Nonlinear  Schr\"{o}dinger equation (NLS)
$$
i \psi^{\bullet}= -\psi'' + 2 |\psi |^2 \psi
$$ 
with periodic boundary conditions.
It has an infinite series of conserved quantities -- integrals of motion $H_1,\;H_2,\cdots$. 
The first three are "classical" integrals 
$$
\aligned 
H_1 &= {1\over 2} \int |\psi|^2 dx={\Cal N}=\text{number of particles},\\
H_2 &= {1\over 2i} \int \psi' \bar{\psi}  dx={\Cal P}=\text{momentum},\\
H_3 &= {1\over 2} \int |\psi'|^2 +|\psi|^4 \, dx ={\Cal H}=\text{energy}.
\endaligned
$$
The others $H_4,\; H_5,\cdots$ do not have  classical names.

In order  to express an invariant Gibbs State 
$$
e^{-H}\prod\limits_{x}d^{\infty}\psi(x) d^{\infty}\bar{\psi}(x)
$$
in action-angle variables, \cite{V1}, one has to write $H$ as a sum of actions
$$
H=\sum\limits_{n}I_n.
$$
The variables  $I$'s  depended, of course,  on the integral $H$. Such formulas for the NLS 
equation were obtain in \cite{MCV1}. The  $I$'s are the integrals of the various
meromorphic 1-forms over the real ovals on the hyperelliptic curve associated with 
the spectral problem for NLS. In this paper we show that $I$'s entering 
into the trace formulas for $H_1$ and $H_3$ and supplemented by some other parameters 
are {\it global moduli} for hyperelliptic  Riemann surfaces with real branching points. 
From another side, they are the {\it actions} in the mechanical sense \cite{A}. 
They are associated with some symplectic structures on the phase space and  we write the
corresponding symplectic forms {\it explicitly}.

As  was demonstrated in \cite{V2} the asymptotics for the ratio of two symplectic volumes
in the limit of infinite genus 
determines the  thermodynamic properties of the Gibbs' state  of completely
integrable system. There was  conjectured a form of the asymptotics for particles
interacting via an elliptic potential. Here we prove formulas of this type for
hyperelliptic  Riemann surfaces which arise in the spectral problem for the NLS equation.

In the papers of  Witten with coauthors, \cite{SW,DW}, the Bogomolny-Prasad-Som\-merfeld  
spectrum of $N=2$ sypersymmetric Yang-Mills theory was expressed in terms of the periods 
of a similar one-form on a Riemann surface. The defining property of the 
{\it meromorphic} 1-form is that the exterior derivative of it is a {\it holomorphic} two-form. 
The subsequent paper of Krichever and Phong, \cite{KP}, provides a firm algebro-geometrical 
framework for  the constructions of physicists.  We essentially use  the results  of \cite{KP} 
in this paper. 

This paper is organized as follows. In sections 2 through 5 we derive various auxiliary 
technical facts. The trace formulas are presented in the section 6. In sections 7 through 9 we 
prepare  the basic technical tool--- the identity for variations of the differential of 
quasimomentum entering into the trace formulas.  The 
variables $I$'s  are studied as moduli in section 
10. We write explicitly  the corresponding symplectic forms in sections 11 and 12. 
In  section 14 we compute the ratio of two symplectic volumes in the infinite genus limit.

\subhead 2. Zero curvature representation \endsubhead
Throughout the paper  the complex form of the Nonlinear Schr\"{o}dinger (NLS) equation 
\footnote"*"{$^\bullet$ denotes a
derivative in the time variable $t$, $'$  in the $x$ variable.}
$$
i \psi^{\bullet}= -\psi'' + 2 |\psi |^2 \psi,
$$
where $\psi= Q+iP$ is replaced by  the more convenient  real form
$$\aligned 
Q^{\bullet}&= -P''+2(Q^2+P^2)P,\\
P^{\bullet}&= \phantom{-} Q''-2(Q^2+P^2)Q. \endaligned
$$
The components $Q(x,t)$,  $P(x,t)$ are infinitely smooth real
functions of  the space variable $x$, periodic with the period $2l_x$.

The equation can be written as a Hamiltonian system
$$
Q^{\bullet}= \{Q,{\Cal H}\}, \quad\quad\quad P^{\bullet}= \{P,{\Cal H}\},
$$
with the Hamiltonian ${\Cal H}=1/2 \int_{-l_x}^{l_x} Q^{\prime2} +
P^{\prime2} + (Q^2+P^2)^2=${\it energy} and the ``classical'' bracket
$$
\{A,B\}=\int_{-l_x}^{l_x} {\partial A\over \partial Q(x)}{\partial B\over \partial P(x)}-
 {\partial A\over \partial P(x)}{\partial B\over \partial Q(x)}\, dx.
$$
The equation has two other ``classical'' integrals of motion
$
{\Cal N}=1/2 \int_{-l_x}^{l_x} Q^2 +P^2=\#${\it of particles}
and ${\Cal P}=\int_{-l_x}^{l_x} QP'=${\it momentum}.

The NLS equation is a commutativity condition for  
$2\times 2$ matrix differential operators
$$
[{\partial\over \partial x}- U, {\partial\over \partial t}-V]=0,
$$
where
$$\aligned
U&=-\lt J+ U_0 =-\lt \pmatrix 0& 1\\
                                 -1& 0 \endpmatrix + \pmatrix Q& P \\
                                                            P& -Q \endpmatrix, \\
V&=\lt^2 J -\l U_0 +(Q^2 +P^2) J -J U_0'\\
 &=\lt^2 \pmatrix 0&1\\ -1&0 \endpmatrix -\l 
 \pmatrix Q&P\\ P&-Q\endpmatrix + (Q^2 +P^2)\pmatrix  0&1\\ -1&0 \endpmatrix  
  + \pmatrix -P'& Q'\\ Q'&P'\endpmatrix.  
\endaligned
$$

\subhead 3. Monodromy matrix. Expansion at infinity \endsubhead
The monodromy matrix $M(x,x_0,\l)$ is a $2\times 2$ solution of
$$\aligned
M'(x,x_0,\l)&=[-\lt J +U_{0}(x)] M(x,x_0,\l),\\
M(x,x_{0},\l)& |_{x=x_0}=I, \endaligned
$$
and it is given by the matrix exponent
$$
M(x,x_0,\l)=\exp \int_{x_0}^{x} U(y,\l)\, dy.
$$
For $QP\equiv 0$ the matrix $M$ can be easily computed
$$ \aligned
M_0(x,x_0,\l)&=e^{-\ltx J}=R\left(\ltx\right)\\
&=\pmatrix \cos\ltx & -\sin\ltx\\
         \sin\ltx & \phantom{-}\cos\ltx  \endpmatrix .
\endaligned
$$
For $QP\neq 0$ and large $\l$ the matrix $M$ is a perturbation of this 
trivial case. It is a solution of the integral equation
$$
M(x,x_0,\l)-\int_{x_{0}}^{x}R\left(\l (x-s)/2\right)U_0(s) 
M(s,x_0,\l)\,ds= R\(\ltx\),
$$
or in symbolic form $[I - A]M=R$. 

Denote for any matrix
$
S=  \pmatrix a & b \\
              c & d  \endpmatrix $ 
two traces $\tr S \equiv a+d,\, \tr_{a}S\equiv b+c.$
\proclaim{Lemma 1} The matrix $M $ can be represented by the convergent series
$$
M=[I-A]^{-1}R=\sum^{\infty}_{n=0}A^nR=M_0+M_1+M_2+\cdots.
$$
The first few terms are listed below
$$\aligned
M_0(x,x_0,\l)=& R\(\ltx \) ,\\
M_1(x,x_0,\l)=& R\(\lxt \)\[-{J\over \l} U_0(x) + {1\over \l^2} U_0'(x)\]\\
             &-  R\(\ltx \)\[-{J\over \l} U_0(x_0) + {1\over \l^2} U_0'(x_0)\],\\
M_2(x,x_0,\l)=& {J\over \l}R\(\ltx \)\int_{x_0}^{x}Q^2+P^2 \\
      &+{1\over \l^2}R\(\ltx \)\left[I\int_{x_0}^{x}(QQ'+PP') + J\int_{x_0}^{x}(QP'-PQ')\right]\\
& -R\(\lxt \){1\over \l^2}U_0(x)U_0(x_0)+ R\(\ltx \){1\over \l^2}U_0^2(x_0).
\endaligned
$$
These are all up to the error\footnote"*"{$\quad|\quad|$ stays for any multiplicative matrix norm.}
$$
|\quad|\leq O\({e^{\Im \ltx}\over \l^{5/2}}\).
$$
The  whole tail $\sum ^{\infty}_{n=3}$ can be estimated as
$$
|\quad|\leq o\({e^{\Im \ltx}\over \l}\).
$$
Moreover, $\tr M_n =0$ for $n$ odd  and $\tr_{a}M_n =0$ for $n$ even.

\endproclaim
\demo\nofrills{Proof.\usualspace} The expression for each term $M_1,\; M_2$ 
{\it etc.}, 
can be obtained from the formula $M_n=A\, M_{n-1}$ 
by integrating by parts. The estimate for the error 
is straightforward from this. The estimate for the whole tail is proved in \cite{MCV1}. To prove 
the statement about traces we need an explicit expression for $M_n$:
$$
M_n(x,x_0,\l)=\int\limits_{x\geq x_1 \geq \cdots \geq x_n \geq x_0} d^nx\, R \, 
U_0(x_1) \cdots U_0(x_n).
$$
The matrix $U_0$ can be written in the form 
$$
U_0(x)=\sqrt{Q^2+P^2}(x) R(\varphi(x)) A,
$$ 
where $A=\(\matrix 0 & 1 \\
                   1 & 0 
                        \endmatrix \)$ and $\varphi(x)$ is some real function. Therefore 
$$
M_n(x,x_0,\l)\int\limits_{x\geq x_1 \geq \cdots \geq x_n \geq x_0} d^nx\,
\prod\limits_{k=1}^{n} \sqrt{Q^2 +P^2}(x_k) R' A^n
$$ 
with some $R'$. Therefore for $n$ even $\tr_{a}M^n=\int d^n x \prod \sqrt{Q^2+P^2} \tr_{a} R'=0$. 
For n odd $\tr M^n=\int d^n x \prod \sqrt{Q^2+P^2}\tr  R' A =0$. 
\qed
\enddemo

\subhead 4.  Spectrum. Some products\endsubhead

The discriminant is defined as $\Delta(\l)={1\over 2} \tr M(l_x,-l_x,\l)$. 
The roots of the equation $\Delta(\l)^2-1=0$ are the points of the periodic/anti\-periodic 
spectrum. When $QP\equiv 0$ we have $\Delta_0(\l)=\cos \l l_x$ and double eigenvalues 
at the points $\l_n^{\pm}={\pi n\over  l_x}.$
If n is even/odd, then the corresponding $\l_n^{\pm}$ belongs to the  periodic/\linebreak 
antiperiodic spectrum.

For generic $QP$,   formal perturbation arguments show that
$$
\l_n^{\pm}={\pi n\over l_x} \pm 2 |\hat{\psi}(n)|+\cdots,
$$
where 
$$
\hat{\psi}(n)={1\over 2l_x} \int\limits^{+l_x}_{-l_x} \psi(x)e^{-i {\pi n \over l_x}x} dx.
$$
The size of the gap is determined roughly speaking by 
the Fourier coefficients of the potential. In fact, for $QP \in C^{\infty}$ 
we have\footnote"*"{$O\({1\over n^{\infty}}\)$ means 
$O\({1\over n^{k}}\)$ for any $k=1,2,\cdots.$}
$$
\l_n^+-\l_n^-=O\({1\over n^{\infty}}\).
$$

We denote by $\mu_n$ the roots of the equation $m_{12}(\l)=0$ and by 
$\l_n^{\bullet}$ the roots of
$\Delta^{\bullet}(\l)=0$. These roots are always cought in the open gaps 
$[\l_n^{-},\l_n^{+}]$.
We can write Hadamard's products for $\Delta^2-1,\; m_{12}(\l)$ and 
$\Delta^{\bullet}(\l)$ as
$$
\align
\Delta^2-1&=-l_x^2 (\l-\l_0^{-})(\l-\l_0^{+})\, \prod\limits_{n\neq 0}
{(\l-\l_n^{-})(\l-\l_n^{+})\over {\pi^2n^2\over l_x^2}}, \tag1 \\
m_{12}(\l)&=-l_x(\l-\mu_0)\,
\prod\limits_{n\neq 0} {(\l-\mu_n)\over {\pi n\over l_x}},\tag 2 \\
\Delta^{\bullet}(\l)&= -l^2_x  (\l-\l_0^{\bullet})\; \prod\limits_{n\neq 0}
{(\l -\l_n^{\bullet})\over {\pi n\over l_x}}. \tag 3
\endalign
$$

\subhead 5.  Floquet solutions. Differential of quasimomentum\endsubhead
The eigenvalues of the monodromy matrix $M(l_x,-l_x,\l)$ can 
be easily computed. 
They are the roots of the quadratic equation $ w^2 -2\Delta w +1=0,$
and given by the formula $w=\Delta\pm \sqrt{\Delta^2-1}$. They become single-valued on the 
Riemann surface (infinite-genus in the generic case) $\Gamma$:
$$
(\l,y)\in {\Bbb C}^2:\quad  y^2= \Delta^2(\lambda) -1.
$$
We denote by $Q$ a  point $(\l,y)$ on the curve $\Gamma$ and by $P_{\pm}$ the two infinities. 

We  introduce a multivalued function $p(Q)$ on the curve $\Gamma$ as $w(Q)=e^{i p(Q)2l_x}$. 
It is
defined up to ${\pi n\over l_x}$, where $n$ is an integer. Let $\tau_0$ be a 
holomorphic involution
on the curve $\Gamma$ permuting sheets $\tau_0: \; (\l,y)\longrightarrow (\l,-y)$,
then $p(\tau_0 Q)+ p(Q)={\pi n\over l_x}$.

The Floquet solution $e(x,Q)$ is defined as a solution which is an eigenvector  of the matrix 
$M(l_x,-l_x,\l):\quad M(l_x,-l_x,\l) \, e(\lx, Q)= w(Q) \, e(\lx,Q)$, normalized by the 
condition $e_1(-l_x,Q)=1$. It is given by the formula
$$
e(x,Q)=\[\matrix
m_{11}(x,\l)\\
m_{21}(x,\l)
\endmatrix \] + {w(Q)-m_{11}(l_x,\l)\over m_{12}(l_x,\l)}
\[\matrix
m_{12}(x,\l)\\
m_{22}(x,\l)
\endmatrix \],\quad \l=\l(Q). \tag 4 
$$
For $QP\equiv 0$ the Floquet solution is   
$e(x,Q)=e^{\pm i \lt(x+l_x)} 
\[\matrix
\phantom{\pm} 1\\
\mp i
\endmatrix \].
$
If $QP\neq 0$ the function $e(x,Q)$ admits assymptotic expansion at  infinities:
$$
e(x,Q)=e^{+i \lt(x+l_x)} \[\matrix
\sum\limits_{s=0}^{\infty} a_s \l^{-s}\\
\sum\limits_{s=0}^{\infty} c_s \l^{-s}
\endmatrix \], \quad   Q\in (P_{+})\quad  \text{and}  \quad a_0=1,\; c_0=-i;
$$
$$
e(x,Q)=e^{-i \lt(x+l_x)} \[\matrix
\sum\limits_{s=0}^{\infty} b_s \l^{-s}\\
\sum\limits_{s=0}^{\infty} d_s \l^{-s}
\endmatrix \],  \quad Q\in (P_{-})\quad  \text{and}  \quad b_0=1,\; d_0=i.
$$
In order to relate these two, consider
$$
e(x,Q) + e(x,\tau_0 Q)= \[\matrix
m_{11}(x,\l)\\
m_{21}(x,\l)
\endmatrix \] + {m_{22}(l_x,\l)- m_{11}(l_x,\l)\over m_{12}(l_x,\l)}
\[\matrix
m_{12}(x,\l)\\
m_{22}(x,\l)
\endmatrix \],
$$
which is a real function for real $\l$. This implies 
$a_s=\overline{b}_s$ and $c_s=\overline{d}_s$ for all $s=0,1,\cdots$. 

Substituting the expansion for $e(x,Q)$ into $\[\partial_x - U(x,\l)\]e=0$ we obtain the reccurent 
relation
$$
-\[\matrix
a_k'\\
c_k'
\endmatrix \] + U_{0} 
\[\matrix
a_k\\
c_k
\endmatrix \]=\( {i\over 2} I + {1\over 2} J\) 
\[\matrix
a_{k+1}\\
c_{k+1}
\endmatrix \], \quad\quad k=0,1,\cdots. \tag 5 
$$
For $k=0$ the formula  producies $Q-iP={i\over 2} a_1 + {1\over 2} c_1$. 
Note that the matrix $ {i\over 2} I + {1\over 2} J$ is degenerate and the coefficients of the 
assymptotic expansion can not be computed reccursively,  
but they can be computed with the aid of Lemma 1  and  formula (4). For example,
$$
\align
a_1& = -i Q - P + iQ(-\lx) +P(-\lx) - i\int\limits_{-\lx}^{x}(Q^2+P^2), \tag 6 \\ 
c_1& = Q-iP + Q(-\lx) - i P (-\lx) - \int\limits_{-\lx}^{x}(Q^2+P^2).\tag 7 
\endalign
$$
Similarily, one can compute the asymptotic expansion for $p(Q)$ at the infinities
$$
p(\l)= \pm \lt + a^{\pm}_{0} + {a^{\pm}_{1}\over \l} + {a^{\pm}_{2}\over \l^2}\cdots.
$$
Indeed, from Lemma 1 
$$\Delta(\l)=\cos l_x \l +{2 {\Cal N} \sin l_x \l \over \l} + \cdots
$$
and using $\Delta(\l)=\cos 2 l_x p(\l)$ we obtain 
$$
\aligned
a^{\pm}_{0} & = {\pi k_{\pm}\over l_x},\\
a^{\pm}_{1} & = \mp {1\over 2l_x} \int_{-l_x}^{l_x} Q^{2} +P^{2} = \mp {1\over l_x} {\Cal N},\\
a^{\pm}_{2} & = \mp {1\over 2l_x} \int_{-l_x}^{l_x} QP'= \mp {1\over l_x}  {\Cal P},\\
a^{\pm}_{3} & = \mp {1\over 2l_x} \int_{-l_x}^{l_x} Q^{\prime2} +P^{\prime2} + (Q^2+P^2)^2=
\mp {1\over l_x} {\Cal H}, \quad \text{etc.}
\endaligned
$$
The differential $dp$ is of the second kind with double poles at the infinities of the form 
$\pm dp=d\(\lt +O(1)\)$ and zero $a$-periods. All these one can see from the formula
$$
dp= \pm {1\over i 2l_x} d \cosh^{-1} \Delta(\l) = \pm {1\over i 2l_x}{\Delta^{\bullet}(\l)\, d\l\over  \sqrt{\Delta^2-1}}.
$$
The $b$-periods of $dp$ are ${\pi n_b\over l_x}$ 
({\it periodicity condition}). This implies that $w(Q)$ is  single-valued on the curve.

\subhead 6.  Trace formulas. One-gap potentials\endsubhead
As in \cite{MCV1} we have
$$
\align
{\Cal N} & = \sum\limits_{n} I_n, \quad \quad \quad 
I_n={l_x\over 2 \pi i} \int_{a_n} p(\l) d\l ,  \tag 8 \\
{\Cal P} & = \sum\limits_{n}I_n', \quad \quad \quad 
I_n'= {l_x\over 2 \pi i} \int_{a_n} \l p(\l) d\l ,  \tag 9 \\
{\Cal H} & = \sum\limits_{n} I_n'', \quad \quad \quad 
I_n''={l_x\over 2 \pi i} \int_{a_n} \l^2 p(\l) d\l , \quad \text{\it etc.} \tag 10 
\endalign
$$
We use these formulas here to determine  positions of branching points  in the case of 
a one--gap potential.  Let 
$$\alpha={\l_n^- +\l_n^+\over 2},\quad \quad \beta={\l_n^+- \l_n^-\over 2}.
$$ 
\proclaim{Lemma 2} For $\psi_n(x)=Ae^{i{\pi n\over l_x}x}$ we have
\roster
\item"{i.}" $\l_n^{\bullet}=\alpha$,
\item"{ii.}" $\beta=2|A|$,
\item"{iii.}" $\alpha={\pi n\over l_x}$.
\endroster
\endproclaim
\demo\nofrills{Proof.\usualspace} For such potential $\l_k^-=\l_k^+$ if $k\neq n$.  
Using Hadamard's products (1) and (3) 
$$
dp= {1\over i2 l_x} {\Delta^{\bullet}(\l)\, d\l\over  \sqrt{\Delta^2-1}} = 
{i\over 2}\, {(\l-\l_n^{\bullet})\over \sqrt{\beta^2-(\l -\alpha)^2}} \, d\l.
$$
The condition $\int_{a_n} dp =0$ implies $\l_n^{\bullet}=\alpha$ and  $i.$ is proved.

If $Q+iP=Ae^{i{\pi n\over l_x}x}$, then ${\Cal N} =l_x|A|^2$.  The integral in 
the right-hand side of the first trace formula can be computed 
$$
{\Cal N}= {l_x\over 2 \pi i } \int_{a_n} \l\, dp(\l)= {l_x \beta^2\over 4}.
$$
This implies $ii.$

The  trace formula for ${\Cal P}$ can be used similarily to prove $iii$.
\qed
\enddemo
\noindent
{\bf Remark.} One can rescale the zero curvature representation as $\l \rightarrow a \l$, where 
$a$ is an arbitrary real number. With our choice of the scaling the potential with the frequency 
${\pi n\over l_x}$ opens the n-th gap with  middle point at $\alpha={\pi n\over l_x}$.

\subhead 7. Infinite hierarchy of commuting flows. BA function\endsubhead
This and the other two sections are preparatory. We obtain the formula for the 
variation of the differential $dp$ wich will be used later.

Consider a finite genus curve  $\Gamma$ with arbitrary real branching points.
The function $e(t,\zeta,Q)$ has asymptotics at infinities $P_{\pm}$
$$
e(t,\zeta,Q)=e^{\pm i\(\sum\limits_{n=1}^{\infty}a_n\l^{n}t_n+\sum\limits_{n=0}^{g}
a_n'\l^n\zeta_n\)} \times
\[\left(\matrix
1\\
\mp i
\endmatrix \right) +\ol\]
$$
and poles at the points $\gamma_1,\cdots,\gamma_{g+1}$, located on the real ovals.
As it is shown in \cite{Kr1} for any $n=1,2,\cdots$ there exists a matrix  
$U_n(x,\l)$ which depends polynomialy on the parameter $\l$ and such that
$$
\[\partial_{t_n}-U_n\]e(t,\zeta,Q)=0.
$$
We limit ourself to the case when $e$ depends only on the two times, $t_1=x,\; t_2=t$,
$$
e(x,t,\zeta,Q)=e^{\pm i\({\l\over 2}x - {\l^2\over 2}t +\sum\limits_{n=0}^{g}
a_n'\l^n\zeta_n\)} \times
\[\left(\matrix
1\\
\mp i
\endmatrix \right) +\ol \].
$$
In this case we can simply write $U_1=U,\; U_2=V$. The quasiperiodic character  of the solution 
$Q(x,t)$ and $P(x,t)$ has been known for a long time, see {\it e.g.} 
\cite{I}. We need a particular form of it.
\proclaim{Lemma 3} \cite{Kr2} The functions $Q(x,t)$ and $P(x,t)$ are 
quasiperiodic functions 
on the  $g+1$ dimentional real torus and the initial position on the torus is   
determined by  the parameters $\zeta=(\zeta_0,\zeta_1,\cdots,\zeta_{g})=(\zeta_0,\zeta_{1g})$ 
$$
Q(x,t)={\hat Q}(\zeta_0, Ux+Wt +\zeta_{1g})\quad \quad 
P(x,t)={\hat P}(\zeta_0, Ux+Wt +\zeta_{1g}),
$$ 
where ${\hat Q}(z_0,z_1,\cdots, z_g),\; {\hat P}(z_0,z_1,\cdots, z_g)$ are the functions 
with unit periods in  all $z$'s and $U,\, W$ are vectors from $R^{g}$. 
\endproclaim
\demo\nofrills{Proof.\usualspace}
First, we introduce multi-valued functions $p(Q),\;E(Q)$ on the curve  
with singularities at  infinities
$$
p(Q)=\pm {\l\over 2} +   a^{\pm} + \ol, \quad \l=\l(Q),\quad \;\;Q\in(P_{\pm}).
$$
Similar
$$
E(Q)=\mp {\l^2\over 2}  + b^{\pm}  + \ol, \quad \l=\l(Q),\quad \;\;Q\in(P_{\pm}).
$$
The function $e(x,t,\zeta,Q)$   can be written in the form 
$$
e(x,t,\zeta,Q)=C\, A\;  \varPhi (Ux+Wt +\zeta_{1g},Q) \, e^{ip(Q)x+iE(Q)t},
$$
where
$$
C= \(\matrix
1 & 1\\
-i& +i
\endmatrix \), \quad A= \left(\matrix
e^{-ia^+ x - ib^+ t +ia_0'\zeta_0} & 0\\
0              & e^{-ia^-x-ib^-t - ia_0'\zeta_0}
\endmatrix \right),
$$
$$ \quad \varPhi= \left(\matrix
\varphi_+(Ux+Wt +\zeta_{1g},Q)\\
\varphi_-(Ux+Wt +\zeta_{1g},Q)
\endmatrix \right)
$$
and $\varphi_{\pm}(z_1,\cdots,z_g,Q) $ are the functions with unit periods in  
$z$'s and rational singularities in $Q$.  

Second, we  expand $e(x,t,\zeta,Q)$ at infinity similarly to section 5 and obtain explicit 
formula for the solution. 

We present complete arguments only for the genus 0.
The curve is rational and given by the equation (see section 6):
$$
y^2=\beta^2-(\l-\alpha)^2.
$$
The uniformizing parameter $z$ is introduced as
$$
\l=z+{\beta^2\over 4(z-\alpha)},\quad\quad
y=i(z-\alpha)+{\beta^2\over 4i(z-\alpha)}.
$$
It is easy to see $z(\gamma_1)=\alpha + z_0=\alpha+{\beta\over 2}e^{i \delta}$ and the 
sheet map $\tau_{0}$ is $\tau_0 z= \alpha +{\beta^2\over 4(z-\alpha)}$.
On such  a curve one can write the differentials $dp,\; dE$ explicitly
$$
\aligned
dp&={i\over 2}{(\l-\alpha)\over \sqrt{\beta^2-(\l-\alpha)^2}}d\l,\\
dE &=i{(\l-\alpha)^2+ \alpha(\l-\alpha) -\beta^2/2 \over \sqrt{\beta^2-(\l-\alpha)^2}}
d \l.
\endaligned
$$
The functions $p(Q)$ and $E(Q)$ are single-valued on the curve and given by the formulas
$$
\align
p(z)& = {1\over 2} (z -{\beta^2\over 4(z-\alpha)}),\\
E(z)& = -{1\over 2} (z^2+ {\beta^2\over 2}) + {1\over 2} 
\(\({\beta^2\over 4(z-\alpha)}\)^2- {2\alpha \beta\over 4(z-\alpha)}\).
\endalign
$$
The parameters in the expansion of $p(Q)$ and $E(Q)$ are $a^+=b^+ = 0$ and 
$a^-=\alpha$ and $b^-=- (\alpha^2+ {\beta^2\over 2})$.

The functions $\varphi$ are given by the formulas $ \varphi_{\pm}=h_{\pm}s_{\pm},$ where
$$
h_+(z)={z-\alpha\over z-\alpha-z_0}, \quad \quad h_{-}(z)=-{z_0\over z-\alpha-z_0};
$$
and $ s_{\pm}(x,t,z)\equiv 1$.  Expanding  $e(x,t,\zeta_0,Q)$ at the infinity $P_+$, say, 
we obtain  the formula for the solution
$$
Q+iP=i{\beta\over 2} e^{-i\delta}e^{i2a_0'\zeta}e^{ +ix\alpha -i\(\alpha^2+{\beta^2\over 2}\)t}.
$$
Choosing suitably the parameter $a_0'$ we prove the statement of the theorem. 

The case of genus $>0$ is considered in \cite{Kr2}.
\qed
\enddemo

\subhead 8. Dual Baker-Akhiezer function \endsubhead
In this section we introduce  the dual BA function. The differential $dp$ has $2g+2$ 
zeros on the real ovals; $g+1$ of them $\gamma_1,\cdots,\gamma_{g+1}$ are related to the other 
$\gamma_1^{\dag},\cdots,\gamma_{g+1}^{\dag}$ by the involution permuting the sheets
$\tau_0 \gamma=\gamma^{\dag}$. 
Now, place the poles of BA functions at the points $\gamma_1,\cdots,\gamma_{g+1}$  and 
define the function $e^{\dag}(x,t,\zeta,Q)$ as\footnote"*"{The sign $+$ here stays for 
a transpose.}
$$
e^{\dag}(x,t,\zeta,Q)=\[e(x,t,\zeta,\tau_0 Q)\]^{+}. 
$$
The  BA functions satisfy the equations
\footnote"**"{The action of the differential operator $D=\sum\limits_{j=0}^{k} \omega_{j} 
\partial^{j}$ 
on the row vector $f^{\dag}$ is defined as 
$f^\dag D=\sum\limits_{j=0}^{k} (-\partial)^{j}(f^{\dag} \omega_j).$}
$$
\[J\partial_{x}-J U\]e(x,t,\zeta,Q)=0,\quad \quad \quad 
e^{\dag}(x,t,\zeta, Q)\[J\partial_{x}-J U\]=0
$$
and similarly in the $t$ variable.
The proof is  by  direct computation and the Riemann-Roch theorem. From the 
results of the previous section 
$$
e^{\dag}(x,t,\zeta,Q)= e^{ip(\tau_0 Q)x+iE(\tau_0 Q)t}\;  \varPhi^{+}\, A\, C^{+}.
$$

The $2\times2$ matrix differential 
$e(t,\zeta,Q)e^{\dag}(t,\zeta, Q)dp (Q)$
has singularities (at most) of  rational character at the infinities. 
The sum of residues for each 
entry must be equal to zero. This requires something hidden from the potential.
Namely, in the periodic case using formulas (2) and (3) we compute  
$$
P(-l_x)=0. \tag 11
$$
This fact will be used later.

\subhead 9. Variational identity \endsubhead
To follow the scheme developed in \cite{Kr2} for 2+1 systems, we introduce an additional 
variable $\epsilon$ such that 
$$
e(\epsilon, x,t,\zeta,Q)=
e(x,t,\zeta,Q)e^{\epsilon \l(Q)},\quad\quad
e^{\dag}(\epsilon, x,t,\zeta,Q)=
e^{-\epsilon \l(Q)}e^{\dag}(x,t,\zeta,Q).
$$
The matrix polynomial $U_k=\sum\limits_{n=0}^{k} u_n^{k} \l^n$ is replaced by the  
differential operator ${\Bbb U}_k=\sum\limits_{n=0}^{k} u_n^{k} \partial^n_{\epsilon}$.
The  "new" BA function satisfies 
$$
\[J\partial_{x}-J {\Bbb U}\]e(\epsilon,x,t,\zeta,Q)=0.
$$ 

Let parameters $I_1,\cdots, I_N$ parametrize the set of spectral curves (not necesarily 
periodic) and $\zeta_0,\cdots,
\zeta_g$ determine the point  on the Jacobian (invariant manifold). 
For such  curves  we denote averaging in the x variable as 
$$
< \bullet>_x= \lim\limits_{L\rightarrow \infty} {1\over 2L} \int_{-L}^{L}dx'\;  \bullet .
$$
In the periodic case it produces the same result as the averaging over the period.
Assume that all $I$'s,  the point $Q$ on the curve depend on the parameter $\tau$ and so are 
$p(Q),\; a^{\pm}$,  {\it etc.} 
Note that the variables $\zeta$'s are fixed.  
We are interested in the variation of $p(Q)$  
in response to the variation of $\tau$.

\proclaim{Lemma 4}\cite{Kr2} The following relation holds 
$$
\align
i \partial_\tau & p(Q) \; <e^{\dag}J e>_{x}-i < e^{\dag} J C \pmatrix \partial_{\tau} a^+ & 0 \\
 0 & \partial_{\tau} a^- \endpmatrix 
A C^{-1} e >_x \\
&+ <  e^{\dag} J C A  \sum\limits_{j=1}^{g} \partial_{\tau} U_{j} {\partial \varPhi\over \partial 
\zeta_{j}} e^{ip(Q) + i E(Q) t } >_x \\
&= - \partial_{\tau} \l(Q) 
<e^{\dag} J \U^{(1)} \,  e>_{x}+ <e^{\dag} J \partial_{\tau} \U \,  e>_{x}.
\endalign
$$
\endproclaim
\demo\nofrills{Proof.\usualspace}
Let $e^{\dag}=e^{\dag}(\epsilon, x,t,\zeta,Q(\tau)),\;  e_1=e(\epsilon, x,t,\zeta,Q(\tau_1)),\;  
U=U(\tau),\; V=V(\tau),\; U_1=U(\tau_1),\; V_1=V(\tau_1)$. They satisfy the identity
\footnote"*"{ we use the formula $(f^{\dag} D)g=\sum\limits_{j=0}^{k} \partial_{\epsilon}^j 
(f^{\dag}(D^{(j)} g)),$ where 
$D^{(0)}=D,\; D^{(1)}=-\sum\limits_{i=1}^{k}j w_j \partial _{\epsilon} ^{j-1}$, {\it etc.}}
$$
\partial_{x}\[e^{\dag}J e_1\]
= - \partial_{\epsilon} \[e^{\dag} J \U^{(1)}  e_1\] +
e^{\dag}\[J {\Bbb U_1}- J {\Bbb U}\] e_1.
$$
Differentiating with respect to $\tau_1$ and assuming $\tau_1=\tau$
$$
\align
 i\partial_{\tau} & p(Q) \; \[e^{\dag}J e\]- 
i [ e^{\dag} J C \pmatrix \partial_{\tau} a^+ & 0 \\
 0 & \partial_{\tau} a^- \endpmatrix
A C^{-1} e ] \\
&+ [  e^{\dag} J C A  \sum\limits_{j=1}^{g} \partial_{\tau} U_{j} {\partial \varPhi\over \partial
\zeta_{j}} e^{ip(Q) + i E(Q) t }]  \\
& = -  \partial_{\tau} \l(Q) \[e^{\dag}  J \U^{(1)} e\] +
e^{\dag}\[J \partial_{\tau} {\Bbb U}(\tau)\]e+ 
R,
\endalign
$$
where $R$ is the remainder term. Let $T_1(\zeta)$ is a subtorus of $T^{g+1}$ which contains 
the closure of the trajectory under $x$-dynamics. Integrate the previous identity with respect 
to the variable $\zeta$ over the flat measure on $T_1(\zeta)$. 
The error term vanishes. Due to the 
ergodicity of the $x$-dynamics on $T_1(\zeta)$ 
one  can replace integration in $\zeta$ by $x$-averaging.
The lemma is proved. 
\qed
\enddemo

\subhead 10. Periods are moduli\endsubhead Consider a 
deformation of the original curve $\Gamma$. 
Namely, assume that  the branch points are differentiable functions of the parameter $\tau$:
$$
\Gamma(\tau): y^2=-\prod\limits_{k=-N}^{k=N}(\l-\l_k^+(\tau))(\l-\l_k^-(\tau)),
\quad\quad\text{where}\quad  
\Gamma(\tau)|_{\tau=0}=\Gamma.
$$
On the curve we have a multivalued function $p(\l,\tau)$, with the asymptotics at infinities 
$$
p(\lambda,\tau)=\pm {\lambda\over 2} + a_0^{\pm}(\tau) + O\({1\over \l}\),
\quad\quad \l=\l(Q),\quad Q\in (P_{\pm})
$$
and normalized so that $a_0^{+}(\tau)=0$. 

We will show that the actions entering into the trace formula (8) and supplimented by the 
$b$-periods are {\it global} coordinates for such curves with real branching points. 
In \cite{MCV1} it
is proved under a periodicity assumption on the curve by direct computation of the Jacobian.
The {\it local} version of such statement is presented in \cite{KP}. 
Here we restrict the class of curves in consideration and therefore are 
able to obtain  a global result. 
Our proof also can be easily generalized for the actions entering into the trace formula (10).
\proclaim{Lemma 5} The map from $\l_{-N}^{-}<\l_{-N}^{+}< \cdots 
< \l_N^{-}<\l_{N}^{+}$
to 
$$
I_1,\cdots, I_{g+1}, U_1,\cdots,U_g,\; a_0^{-}(\tau),
$$ 
where
$$
I_k(\tau)={l_x\over 2\pi i} \int_{a_k} p (\lambda,\tau)d\lambda, \quad \quad k=1,2,\cdots,g+1;
$$ 
$$
U_k(\tau)=\int_{b_k}d p(\lambda,\tau), \quad \quad k=1,2,\cdots,g.
$$
has  nonvanishing Jacobian.
\endproclaim
\demo\nofrills{Proof.\usualspace} Consider a   deformation, such that 
$
\partial_{\tau}I_k(\tau)|_{\tau=0}=\partial_\tau U_k(\tau)|_{\tau=0}=0
$
for all $k$ and also $\partial_\tau a_0^{-}(\tau)|_{\tau=0}=0$.
This is equivalent to saying that the response of these parameters  to the variation of 
$\tau$ is of  order $O(\tau^2)$. 
Outside of  the branching points the function $p(\lambda,\tau)$ is  defined up 
to\footnote"*"{$a$-periods of the differential $dp$ are zero.}
$$
\sum_{k=1}^g n_kb_k\left[d\, p(\lambda,0)\right]+O(\tau^2),
$$
where $n_k$ are some integer numbers.  Therefore,  the derivative 
$\partial_\tau p (\lambda,\tau)$ is a single-valued function on the curve. 

In the vicinity of any branch point, say $Q_k^{+}=(\l_k^{+},0)$ the 
differential $dp(\lambda,\tau)$ can be represented by the series
$$
dp(\lambda,\tau)=\sum\limits_{k=0}^{\infty} {\omega_k\over k!} 
(\l-\l_k^{+}(\tau))^{k- {1\over 2}} \, d\l.
$$
Then 
$$
p(\lambda,\tau)=\int^{\l}dp(\lambda,\tau)= 
\sum\limits_{k=0}^{\infty}{\omega_k\over k!} {1\over k+{1\over 2}}(\l-\l_k^+)^{k+{1\over 2}} 
+ \text{regular part}.
$$
Therefore 
$$
\partial_\tau p(\lambda,\tau)= - \omega_0 (\l-\l_k^+)^{-{1\over 2}} \partial_{\tau} \l^+_k + 
\text{regular part},
$$ 
and
$$
\text{res} \; \partial_\tau p (\lambda,\tau) =-\omega_0 \partial_\tau \l^+_k (\tau). 
$$

The rest of the proof consists of two parts. First, we show that $\partial_{\tau} p\equiv 0$. 
Second, we show that $\omega_0\neq 0$ for all branch points. These lead to $\partial _{\tau} \l_k^
{\pm}  =0$ and prove the result.

The condition $\partial_\tau I_k(\tau)|_{\tau=0}=0$ implies 
$$
\int\limits_{Q_k^-}^{Q_k^+}  \partial_{\tau} p \, d\l =0 \quad \quad k=1,\cdots,g+1. \tag 12
$$ 
Therefore $\partial_{\tau} p$ has at least two zeros on each real 
oval. The condition $\partial_{\tau} a^{-}_0=0$ makes $\partial_{\tau} p$  vanish at the
infinities $P_{\pm}$.  Therefore $\partial_{\tau} p$  has $2g+2$ poles at the 
branching points, $2g+4$ zeros and vanishes identically.
 
In order to show that zeros of $d p$ never match branch points 
$Q_k^{\pm}$ we use the variational identity of Lemma 4. We have $\partial_{\tau} \U= 
\partial_{\tau} a_0^{\pm}= 
\partial_{\tau} U_j =0,\quad j=1,\cdots, g$ and $\U^{(1)} ={1\over 2} J$. Therefore 
$$
i \partial_{\tau}p(Q) <e^{\dag}(x,Q)Je(x,Q)>_{x}={1\over 2} \partial_{\tau} \l(Q)
<e^{\dag}(x,Q)e(x,Q)>_{x}
$$
The points $Q_k^{\pm}$ are fixed points of the 
involution $\tau_0$ and $e^{\dag}(x,Q)=e^{+}(x,Q)$ there. Moreover
$$
<e^{\dag}(x,Q)e(x,Q)>_{x}>0\quad\text{and}\quad <e^{\dag}(x,Q)Je(x,Q)>_{x}=0 \quad \text{at}\quad
Q=Q_k^{\pm}.
$$
This implies that $dp (Q_k^{\pm})\ne 0$. The lemma is proved.
\qed
\enddemo

\noindent
{\bf Remark.} The same statement is true for the actions entering into the trace formula (10), 
supplimented by periodicity conditions and $a_0^-$. The proof is exactly the same with 
identity (12) replaced by
$$
\int\limits_{Q_k^-}^{Q_k^+} \l^2  \partial_{\tau} p \, d\l=0 \quad \quad k=1,\cdots,g+1. 
$$

\subhead 11. Classical symplectic structure\endsubhead
In this section we express the algebro-geometrical 2-form 
$$
\omega=\sum\limits_{k=1}^{g+1} i \delta p(\gamma_k) \wedge \delta \l(\gamma_k)
$$
entering into the trace formula (8) in $QP$ coordinates. 
The following identity, \cite{KP},  holds
$$
\omega= \sum\limits_{P_{\pm}} \text{res}\; 2i {<\delta e^{\dag} \wedge J \delta U_0 e>_x\over 
<e^{\dag} e>_x} dp .
$$
The proof can be obtained folowing  \cite{KP} with 
the use of variational identity of Lemma 4. 
It is enough to compute contribution at one of the infinities, say at $P_+$. 
At another infinity, $P_-$,  it is the same.

From the definition
$$
\align
&e(x,Q)=e^{+ i\lt(x+l_x)} \[\pmatrix 1\\ -i \endpmatrix + z \pmatrix a_1\\ c_1 \endpmatrix 
+\cdots \], \\
&\delta e^{\dag}(x,Q)= e^{- i\lt(x+l_x)} \[ z (\delta \bar{a}_1, \delta \bar{c}_1) +\cdots \],\\
&\delta U_0= \pmatrix \delta Q & \delta P \\
                   \delta P & -\delta Q \endpmatrix.
\endalign
$$
Computing the residue
$$
{1\over 2} \omega= < (\delta \bar{a}_1, \delta \bar{c}_1)\wedge 
\pmatrix \delta P +i \delta Q \\ \delta Q +i \delta P \endpmatrix>_x.
$$
Using the formula $Q-iP= {i\over 2} a_1 + {1\over 2} c_1$ we finally obtain
$$
\omega= 4  <\delta Q \wedge \delta P>_x.
$$
{\bf Remark.} This result was obtain in a different way in 
\cite{MCV2}, see also \cite{MC}.

\subhead 13. Higher symplectic structures\endsubhead
Now we can write in $QP$ coordinates the form 
$$
\omega'=\sum\limits_{k=1}^{g+1} i \delta p(\gamma_k) \wedge \delta \l^{2}(\gamma_k).
$$
entering into the trace formula (9).
Using the identity
$$
\omega'= \sum\limits_{P_{\pm}} \text{res}\; 2i {<\delta e^{\dag} \wedge J \delta U_0 e>_x\over
<e^{\dag} e>_x} \l(Q)dp 
$$
similar to the previous section we compute at $P_{+}$
$$
{1\over 2} \omega'= i< (\delta \bar{a}_1, \delta \bar{c}_1)\wedge 
\pmatrix \delta P a_1- \delta Q c_1\\ -\delta Q a_1 - \delta P c_1 \endpmatrix +
(\delta \bar{a}_2, \delta \bar{c}_2)\wedge
\pmatrix \delta P + i\delta Q \\ -\delta Q + i\delta P  \endpmatrix>_x.
$$
For $k=1$ the formula (5) produces 
$$
a_2-ic_2=Q'-iP' + X (Q-iP), \quad \quad \text{where} \;\; X=-ia_1+ c_1.
$$
This leads to
$$
{1\over 2} \omega'= <\delta Q'\wedge \delta Q +\delta P'\wedge \delta P  +
\delta \bar{X} \wedge (\delta Q^2+ \delta P^2) -2i (\bar{X}-X) \delta Q \wedge\delta P>_x.
$$
Now using the formulas (6-7) and (11) we finally obtain
$$
\omega'= 2<\delta Q'\wedge \delta Q +\delta P'\wedge \delta P  + (\delta Q^2+ \delta P^2)
\wedge \partial^{-1} (\delta Q^2+ \delta P^2) >_x,
$$
restricted to  submanifold
$$
\int\limits_{-l_x}^{l_x} Q^2+P^2 =const. \tag 13
$$
The constant of integration in $\partial^{-1}=\int_{-l_x}^{x}  dx' $ is irrelevant 
since 1-form $<Q\delta Q + P\delta P>_x$ vanishes on the vector fields tangent to the 
sphere (13).

\noindent
{\bf Remark.} This form of the second symplectic structure was conjectured in \cite{MCV1}.

\subhead 14. Symplectic volume elements\endsubhead
We start this section with a simple example of the linear Schr\"{o}dinger equation 
with $2l_x$-periodic  potential
$$\aligned
Q^{\bullet}&= -P'',\\
P^{\bullet}&= \phantom{-} Q''. \endaligned
$$
It has three classical integrals of motion, exactly like in the cubic case
$$
\align
{\Cal N}& ={1\over 2} \int_{-l_x}^{l_x} Q^2 +P^2= \sum\limits_{n} I_n, \quad \text{with} \quad 
I_n = l_x |\hat{\psi}(n)|^2, \\
{\Cal P}& =\int_{-l_x}^{l_x} QP'=\sum\limits_{n} I_n', \quad \text{with} \quad              
I_n' = l_x \({\pi n\over l_x}\)  |\hat{\psi}(n)|^2, \\
{\Cal H}& ={1\over 2} \int_{-l_x}^{l_x} Q^{\prime2} +
P^{\prime2} = \sum\limits_{n} I_n'' , \quad \text{with} \quad              
I_n'' =  l_x \({\pi n\over l_x}\)^2 |\hat{\psi}(n)|^2.
\endalign
$$
The variables $I_n, I_n' $ and $ I_n''$ are actions but relative to the different symplectic 
structures
$$
\align
\omega&= \int\limits_{-l_x}^{l_x} dx\, dQ(x) \wedge d P (x),  \\
\omega'&= \int\limits_{-l_x}^{l_x} dx\, dQ'(x)\wedge dQ(x) + d P '(x)\wedge P (x),  \\
\omega''&= \int\limits_{-l_x}^{l_x} dx\, dQ'(x)\wedge P '(x). 
\endalign
$$
All the actions are conjugate to the same angles $\varphi_n=\text{phase}\; \hat{\psi}(n)$ in the 
corresponding bracket. 

Now consider  $\omega_N,$ {\it etc.} the restriction of the forms on the subspace 
spanned by the first $2N$ harmonics $\psi_n(x)= e^{i{\pi n\over l_x}x},\; n=\pm 1,\cdots,\pm N$. 
Using 
$$
\omega= dI_{-N} \wedge d \varphi_{-N} +\cdots + dI_{N}\wedge d \varphi_{N},\quad \text{\it etc.} 
$$
we obtain in the limit of infinite dimension
$$
\align
{1\over 2N} \log & {\overset 2N\to\wedge \omega_N'\over  
\overset 2N\to\wedge \omega_N}=  {1\over 2N} \log \prod \Sb -N \leq n \leq N\\ n\ne 0 \endSb   
{\pi n\over l_x}\\ 
&=\log N +  \[\log {\pi \over  l_x} - 1\] + {\log 2\pi N\over 2N}+ o\({1\over N}\),\\
{1\over 2N} \log & {\overset 2N\to\wedge \omega_N''\over \overset 2N\to\wedge \omega_N}=
{1\over 2N} \log\prod \Sb -N \leq n \leq N \\ n\ne 0 \endSb   \({\pi n\over l_x}\)^2\\
& = 2 \log N + 2 \[\log {\pi \over  l_x} - 1\] + {\log 2\pi N\over N}+ o\({1\over N}\).
\endalign
$$

The picture in the case of cubic NLS is more complicated, but the shape of the formulas is 
similar.

Consider a submanifold in the 
function space with $2N+1$ gaps open and all other closed. Denote as before 
by $\omega_N,\; \omega_N'$ 
the restriction of the symplectic forms $\omega,\; \omega'$ on this submanifold.

\proclaim{Lemma 7} In the limit of infinite genus the  asymptotic identity holds
$$
\align
{1\over 2N} \log & {\overset 2N\to\wedge \omega_N'\over \overset 2N\to\wedge \omega_N}= 
\log N + \[ \log {\pi \over  l_x} - 1\] + {\log 2\pi N\over 2N}+ \\ 
&+{1\over 2N}\[ \log \int\limits_{[0,1)^{\infty}} m_{12}(0)\ d^{\infty} \tilde{\theta} 
-  \log l_x\] +o\({1\over N}\).
\endalign
$$
\endproclaim
\demo\nofrills{Proof.\usualspace} The strategy of the proof is similar to \cite{MCV2}.
The points of the divisor $\gamma$'s can be considered as a coordinates on the invariant 
torus. The canonical pairing implies 
$$
\det \[{\partial I'_i \over \partial I_j}\] = \det \[{\partial \theta_i \over \partial 
\theta'_j}\] =
{\det \[ \omega_i/d\l(\gamma_k)\]\over \det \[\omega'_j/d\l(\gamma_k)\]},
$$
where the angles $\theta, \; \theta'$ are given by the Abel sum
$$
\theta_i=\sum  \Sb -N \leq k \leq N\\ n \ne 0 \endSb \int\limits_{Q^-_k}^{\gamma_k} \omega_i,
\quad \quad \quad 
\theta_i'=\sum  \Sb -N \leq k \leq N\\ n \ne 0 \endSb \int\limits_{Q^-_k}^{\gamma_k} \omega_i'
$$
where
$$
\omega_i=\sum\limits_{j=1}^{2N} c_{ij}{\l^{j-1}\over R} d\l, \quad \quad \quad
\omega_i'=\sum\limits_{j=1}^{2N} c_{ij}'{\l^{j-1}\over R} d\l
$$
are normalized diferentials of the third kind.
Using normalization $a_k[\omega_j]=a_k[\l\omega_j']=\delta_{kj}$ we obtain
$$
\det \[{\partial I_i'\over \partial I_j}\]= {1\over l_x} 
\(\prod\limits_{ 1 \leq k \leq N} 
{\pi n\over l_x}\)^2
\int\limits_{[0,1)^{2N+1}} {\prod\limits_{-N \leq k \leq N} \mu_k\over {1\over l_x}
\(\prod\limits_{ 1 \leq k \leq N}
{\pi n\over l_x}\)^2} d^{2N+1} \tilde{\theta}_k,
$$
where $\tilde{\theta}_k=\theta_k/ 2\pi$ are normalized angles. Now, using the formula (2) 
we recognize under the integral  $m_{12}(0)$ averaged over the flat 
measure on the invariant torus.  Taking the logarithm of both parts we obtain the statement.
\qed
\enddemo

\noindent
{\bf Acknowledgments.} Finally, I would like to thank I.M. Krichever for 
numerous stimulating discussions. Also 
I am grateful to L. Craine for reading the manuscript and suggesting improvements.

\bye